\documentclass[aps,prc,showpacs,twocolumn,preprintnumbers,floatfix,
showkeys,tightenlines]{revtex4}
\usepackage{amsmath,amssymb,amsfonts}
\usepackage{graphicx}
\usepackage{color}
\allowdisplaybreaks

\newcommand{\be}{\begin{equation}}
\newcommand{\ee}{\end{equation}}
\newcommand{\bea}{\begin{eqnarray}}
\newcommand{\eea}{\end{eqnarray}}
\newcommand{\bm}{\bibitem}

\newcommand{\ep}{\epsilon}

\newcommand{\ze}{\zeta}

\newcommand{\cz}{{\cal Z}}

\begin{document}

\setcounter{page}{1}

\vspace*{0.5 true in}

\title{The efficacy of isotope thermometry: examining in the $S$-matrix
approach}

\author{S.K. \surname{Samaddar}}
\email{santosh.samaddar@saha.ac.in}
\author{J.N. \surname{De}}
\email{jn.de@saha.ac.in}
\affiliation{Theory Division, Saha Institute of Nuclear Physics, 1/AF 
 Bidhannagar, Kolkata 700064, India.}  

\date{\today}

\begin{abstract} 
Isotope thermometry, widely used to measure the temperature of a hot
nuclear system formed in energetic nuclear collisions, is examined in
the light of $S$-matrix approach to the nuclear equation of state of
disassembled nuclear matter. Scattering between produced light fragment 
pairs, hitherto neglected, is seen to have an important bearing on the
extraction of system temperature and volume at freeze-out from isotope
thermometry. Taking due care of the scattering effects and decay
of the primary fragments, a more reliable way to extract the nuclear
thermodynamic parameters is suggested by exploiting least-squares fit
to the observed fragment multiplicities.
\end{abstract}


\maketitle
Theoretical investigations on the equation of state (EOS) of infinite
\cite{goo,sto,mul,lee} and finite \cite{gro,bon,de1,das,cho,de2} 
nuclear matter predict the existence 
of a liquid-gas (LG) type phase transition in these systems. This transition
is thought to play an important role in nucleosynthesis in supernova 
explosion \cite{lam,ish}. Laboratory experiments in collisions between 
energetic nuclei appear to reveal signals of LG phase transition in
hot finite nuclear systems \cite{poc,hau,nat}. Proper identification of such
a transition, however, depends on the reliable measurement of the
thermodynamic observables. In particular, temperature plays a pivotal
role. An widely used practice to extract the temperature of hot nuclear
systems is to take resort to double-isotope ratio thermometry as suggested
by Albergo {\it et al} \cite{alb}. In an ideal scenario, the primary 
fragments produced in the freeze-out volume are assumed to be in their
ground states. Particle and $\gamma $-decay corrections to the excited
primary fragments have also been built in \cite{kol}. Generally,
the feeding effect of secondary decay has been accounted through a 
correction factor \cite{tra,poc,sfi,tsa} on the measured multiplicities. The 
temperature is seen to increase by $\sim $ 10-20 \% from 
the ideal situation. 

All these analyzes have been done with the assumption that the
fragment species produced are noninteracting within the freeze-out
volume. Strong interaction corrections, appropriately taken up in
the $S$-matrix approach \cite{dash} to the grand partition function
of the dilute nuclear system, where in addition to all the stable
mass particles, the two-body scattering channels between them can
be included systematically are seen to modify the fragment multiplicities
\cite{hor,mal}. The extracted temperature as obtained in the previous
analyses without strong interaction corrections may then differ
from the real temperature at which the fragments were produced.
In a schematic calculation \cite{sam} in $S$-matrix approach in
dilute infinite nuclear matter, neglecting secondary decay, we
found that the scattering effects on the extracted temperature and volumes
are not negligible. In the present communication, these ideas are 
incorporated to provide a realistic framework to analyze the data 
in an experimental multifragmentation set up to extract the temperature
and volume of a finite disassembling nucleus at freeze-out with
explicit inclusion of $\gamma $ and particle decay as well as 
effects from scattering between different fragment species.

The details of the $S$-matrix approach, as applied to nuclear systems, 
are given in Refs.~ \cite{mal,sam,sam1}. 
For completeness, a few relevant
equations are presented here highlighting the 
approximations. The grand partition function ${\cal Z}$ of a system
in thermodynamic equilibrium 
can be written as a sum of three terms \cite{sam}  

\begin{eqnarray}
\ln \cz =\ln \cz_{gr}+\ln \cz_{ex}^0 +\ln \cz_{sc}~.
\end{eqnarray}
The first and second terms correspond to the contributions from the
ground states and particle-stable excited states of all the produced
fragment species behaving like an ideal quantum gas.
The last term sums up the contributions
from the scattering states, expressible in terms of the $S$-matrix 
elements. Formal expressions for these three terms are spelt out
in Ref. \cite{mal}.

The scattering channels, for convenience, can be separated into two parts,
one containing only light particles and the other the heavy ones,
{\it i.e.},
\be
\ln {\cz}_{sc} = \ln {\cz}_{sc}^l + \ln {\cz}_{sc}^h.
\ee
The scattering of the heavy ones is dominated by a multitude of
resonances near the threshold; the $S$-matrix elements can then be
approximated by resonances, which like the excited states are again
treated as ideal gas terms \cite{das1}. These are the particle unstable 
states. Structurally, $\ln {\cz}_{sc}^h$ being then similar to
$\ln {\cz}_{ex}^0$, $\ln {\cz}_{ex}^0$ and
$\ln {\cz}_{sc}^h$ are combined together to give $\ln {\cz}_{ex}$
($\equiv \ln {\cz}_{ex}^0+\ln {\cz}_{sc}^h$), which contains
contributions from particle-unstable excited states besides
the particle-stable ones. 

In $\ln {\cz}_{sc}^l$, only the elastic scattering channels for the
pairs $NN, Nt, N{^3}He, N\alpha $ ($N$ refers to the nucleon) and
$\alpha \alpha $ have been included. These calculations involve 
virial coefficients \cite{mal,hua}
that are functions of only
experimental entities, namely, phase shifts and binding energies.
Once the partition function is obtained, total fragment multiplicities
$Y_i$  for the $i$-th fragment species with $N_i$ neutrons
and $Z_i$ protons can be evaluated as
\bea
Y_i =\ze_i\left ( \frac{\partial}{\partial \ze_i}
\ln {\cz}\right )_{V,T}~. 
\eea
Here $\zeta_i [\equiv \zeta_{Z_i,N_i} ]$ is the effective fugacity
defined as $\zeta_{Z_i,N_i} = e^{\beta (\mu_{Z_i,N_i}+B(A_i,Z_i))}$. 
$B(A_i,Z_i)$ is the binding energy of the fragment and $\mu_{Z_i,N_i}$ 
is its chemical potential, which from chemical equilibrium is
$\mu_{Z_i,N_i}=N_i\mu_n+Z_i\mu_p$, $\mu_n$ and $\mu_p$ being the
neutron and proton chemical potentials obtained from the
conservation of the total neutron and proton numbers of the system,
$\beta $ is the inverse temperature.

For relatively low density and not too low temperature, assuming that
the quantal distribution can be replaced by a classical one,
expressions for the primary fragment multiplicities of the $i$th
species can be derived as,
\bea
Y_i&&=V\frac{A_i^{3/2}}{\lambda^3}e^{[\mu_n N_i +\mu_p Z_i +B(A_i,Z_i)]/T}
\nonumber \\
&&\times \left [g_0^i+\sum_{\ep_j =\ep_0}^{\ep_r}
g_j^ie^{-\ep_j^i/T}\right]+Y_{sc}^i~.
\eea
In Eq.~(4), $V$ is the volume of the system, $\lambda = \sqrt {
2\pi/mT } $ (we use natural units $\hbar =c=1$) 
is the nucleon thermal wavelength and $g_0^i$ and
$g_j^i$ are the degeneracies of the ground and excited states.
 The sum over the excited states includes both $\gamma $ and particle-decay
(resonance) channels. In different variants 
of the models of nuclear statistical
equilibrium, only the first term (that also implicitly contains
scattering corrections from resonances in heavy fragments) on the
right hand side of Eq.~(4) has been used to obtain the nuclear 
thermodynamic observables. The last term $Y_{sc}^i$ is
the contribution to the fragment yield from scattering, it
is  nonzero
only for the fragments in the light species set. Expressions for
the multiplicity yields $Y_{sc}^n$, $Y_{sc}^p$ etc. (collectively 
written as $Y_{sc}^l$) are given in Ref. \cite{hor}. From now on, corrections 
obtained with the use of $Y_{sc}^l$ in the extraction of nuclear 
parameters would be called scattering  corrections.

The multiplicities of the primary excited fragments as obtained in 
Eq.~(4) undergo changes because of subsequent particle emission. 
The secondary yield can be written in terms
of the variables $V, \mu_n, \mu_p $ and $T$ at freeze-out as 
follows. For light fragments ($A_i \leq 4, Z_i \leq 2 $),
\bea
Y_i(A_i,Z_i)&&=V g_0^i\frac {A_i^{3/2}}{\lambda ^3}
e^{[(N_i\mu_n+Z_i\mu_p+B(A_i,Z_i))/T]}+ \nonumber \\
&&V\sum_{j}\sum_{k_j} \{\frac{A_j^{3/2}}{\lambda^3}
e^{[(N_j\mu_n+Z_j\mu_p+B(A_j,Z_j))/T]} \nonumber \\
&& \times\omega_p^{k_j}
(A_j,Z_j,T)x_i^{k_j}(A_j,Z_j,T) \} +Y_{sc}^i.
\eea
The light fragments are assumed to be produced only in their ground
states, their multiplicities being given by the sum of the 
first and the last terms in Eq.~(5). 
Their population is further fed from decay of heavier species
given by the second term.
The sum $j$ runs over all species with $A_j >$ 4 and $Z_j\ge $ 2  having
particle-unstable excited states and the sum $k_j$ runs over all the
particle-decaying states of the $j$-th species. The quantity $x_i^{k_j}$
corresponds to the branching ratio of the $ k_j$-th state for emitting the 
$i$-th species; 
it is calculated using the Weisskopf-Ewing
model \cite{wei}. The quantity $\omega_p^{k_j} (A_j,Z_j,T)$ is 
the internal partition function for the particle-unstable states
\bea
\omega_p^{k_j}(A_j,Z_j,T)=g_{k_j}e^{-\ep_{k_j}/T}.
\eea

For heavy particles ($A >4, Z\geq 2$), the observed yield is
\bea
&&Y(A,Z)=V\frac {A^{3/2}}{\lambda^3}e^{(N\mu_n+Z\mu_p+B(A,Z))/T} \nonumber \\
&&\times \{ g_0(A) +\omega_{\gamma}(A,Z,T) \nonumber \\
&&+\sum_{k_j}\sum_{i=1}^6 (\frac {A+a_i}{A})^{3/2} 
e^{(n_i\mu_n+z_i\mu_p+B(A+a_i,Z+z_i)-B(A,Z))/T} \nonumber \\
&&\times\omega_p^{k_j}(A+a_i,Z+z_i,T)x_i^{k_j}(A+a_i,Z+z_i,T)\}.
\eea
In Eq.~(7), $\omega_{\gamma} $
$(=\sum_kg_ke^{-\ep_k/T})$ is the partition function for $\gamma$-decaying 
states, the sum $i$ runs over the emitted ejectiles for which we take
only $n,p,d,t,^{3}$He and $\alpha $, $a_i,z_i$ being their mass and 
charge.  Kolomiets {\it et al. }\cite {kol} arrived also at expressions
of the type given in Eqs.~(5) and (7), the important difference being 
the absence of the scattering correction and consideration of only
the dominant decay mode. They further considered only nucleon and
$\alpha $-decay channels. Moreover, the feeding to the light fragment
yield was neglected. Given a set of experimental yields for four
fragments, their single ratios are constructed using Eqs.~(5) and 
(7) resulting in a system of three independent equations. The equations are
solved iteratively in Newton-Raphson method yielding values of
$\mu_n, \mu_p $ and $T$. The volume can then be determined knowing
the yield of a fragment. 

To explore the effect of scattering on the extracted values of $T$
and $V$ of a hot fragmenting system, we take resort to a numerical 
experiment. The primary fragment yields are calculated with given 
freeze-out temperature $T_{fz}$ and volume $V_{fz}$ in the $S$-matrix
approach as elucidated. The secondary yields are then calculated 
using Eqs.~(5) and (7). These are taken as {\it observed numerical
data}. In Eq.~(7), the first term in the braces corresponds to
the ideal Albergo condition yielding $T_{alb},V_{alb}$, addition of 
the second term gives the $\gamma$-decay corrected values $T_{\gamma},
V_{\gamma}$, further addition of the last term gives $\gamma +p$
(particle)-decay corrected values $T_{\gamma +p}, V_{\gamma +p}$.
Only if heavy fragments are taken for multiplicity ratios, 
$T_{\gamma +p}=T_{fz}$ and $V_{\gamma +p}=V_{fz}$. For light
fragments, to arrive at the actual values of $T_{fz}$ and $V_{fz}$, one has
to further consider scattering corrections 
as given by the last term in Eq.~(5).

\begin{figure}[t]
\includegraphics[width=0.45\textwidth,clip=false]{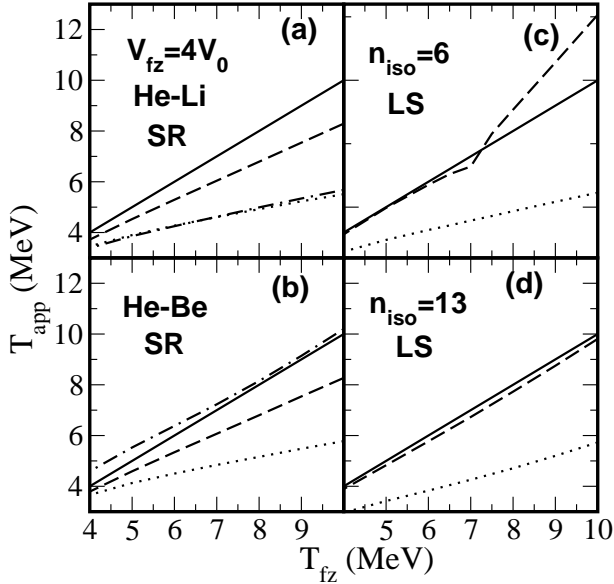}
\caption{In panels (a) and (b), the
extracted temperature $T_{app}$ for the fragmenting system
$^{124}$Sn as a function of the freeze-out temperature $T_{fz}$ shown 
 for two different thermometers
under different approximations using single ratios (SR). In panels (c)
and (d) the same are shown using least-squares method (LS)
for two sets of isotopes as mentioned in the text. The dashed-dot, 
dotted, dashed and full lines correspond to $T_{alb}$, $T_{\gamma}$,
$T_{\gamma +p}$ and $T_{\gamma +p+sc}=T_{fz}$, respectively.}
\end{figure}

\begin{figure}[t]
\includegraphics[width=0.45\textwidth,clip=false]{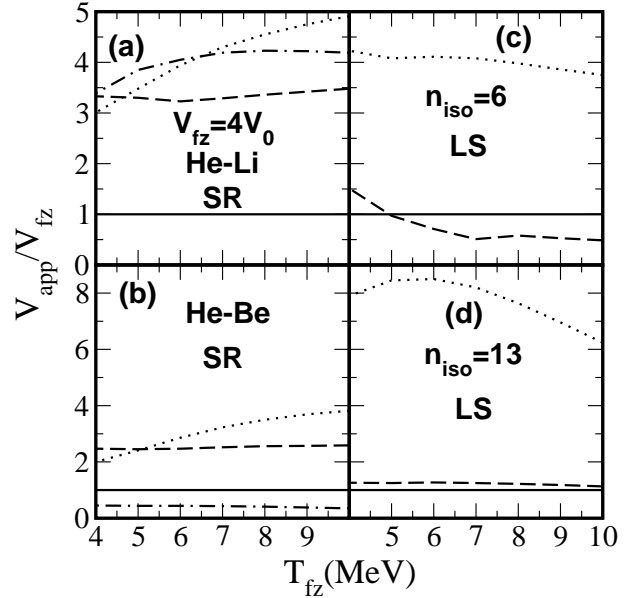}
\caption{The same as in Fig.~1 for the extracted volume $V_{app}$
in units of freeze-out volume $V_{fz}$.}
\end{figure}

The calculations are done with $^{124}$Sn as a representative
system. For the fragment species, all the stable nuclei upto
$A$=124 and $Z$=50 as well as their binding energies are taken
from \cite{mye}. 
All discrete levels upto an excitation energy of 20 MeV with lifetimes
$>$ 200 fm/c as well as their decay modes for $5\leq A \leq 16$ have
been taken \cite{aje,til} into consideration. For still heavier nuclei,
the sum over excited states in Eqs.~(5) and (7) is replaced by an integral
convoluted with the single-particle level density 
$\omega (A,E)$ \cite{boh,mal}.
The integration limits are taken between 2 MeV (approximated for
the first excited state) and  8 MeV (the last particle-stable state)
for the $\gamma$-decaying levels; the resonance limit is taken as
20 MeV. 
The calculations done at different temperatures 
in a freeze-out volume $4V_0$ ($V_0$ is the 
normal volume of $^{124}$Sn) are presented. We have chosen two sets of four
fragments, namely, $^{3}$He, $^{4}$He, $^{6}$Li, $^{7}$Li and
$^{3}$He, $^{4}$He, $^{10}$Be, $^{11}$Be which we refer to as
He-Li and He-Be thermometers. The extracted apparent temperatures $T_{app}$
are found to be quite sensitive to the different approximations
as displayed in the left panels of Fig.~1. 
Except for $T_{alb}$, the other temperatures
are not very sensitive to  the choice of 
thermometer. 
Successive improvement of approximations is seen
to bring the apparent temperature closer to the real one. With inclusion
of effects due to $(\gamma +p)$-decay and scattering, the apparent
temperature $T_{\gamma +p+sc}$ when calculated yields the 
actual temperature $T_{fz}$. The effect of scattering 
is seen to be substantial.

The volume $V_{app}$ (measured in units of $V_{fz}$) extracted in
different approximations is displayed in the
left panels of Fig.~2 as a function of
$T_{fz}$ for the above mentioned  thermometers.
Scattering has a comparatively  more 
significant role here than that observed in
the determination of temperature. Its inclusion collapses the apparent
volumes $V_{app}/V_{fz}$ to unity.

The method so discussed suffers from two limitations. For many thermometers,
there may not be convergence for the solution as noted earlier 
\cite{kol}. We also found that there may be multiple solutions. We
have presented those solutions that are robust in the sense that
taking a considerable range of initial guess values in the iterative method,
same solutions are obtained. To overcome these limitations, we
propose that the least-squares fit to the secondary multiplicities may
be more
fruitful in extracting the temperature and volume. Given experimental
yields for a chosen number of fragment species $n_{iso}$, the 
least-squares fit to 
\begin{eqnarray}
\sum_{i=1}^{n_{iso}}[Y_i^{exp}-Y_i(T,V,\mu_n,
\mu_p)]^2=\chi^2
\end{eqnarray}
 has been performed. The quantities $Y_i^{exp}$
are the  experimental multiplicities which are functions
of the thermodynamic variables at freeze-out and 
$Y_i(T,V,\mu_n,\mu_p)$ are the yields calculated from Eqs.~(5) and
(7) with various approximations as explained earlier. In our
calculations, $Y_i^{exp}$ are taken from our {\it numerical experiment}.
The extracted temperatures $T_{app}$ 
in the least-squares method for the system $^{124}$Sn at 
different given $T_{fz}$ and at a freeze-out volume $V_{fz}=4V_0$ 
under different approximations
are displayed in the right panels of Fig.~1. 
The calculations have been performed using 
a set of light isotopes with $n_{iso}$ =6 ($n,p,d,t,^{3}He$ and $^{4}He$).
The calculations are repeated with a
broader set ($n_{iso}$=13) that includes, besides the light set also the
nuclei $^{6}$Li, $^{7}$Li, $^{9}$Be, $^{10}$B, $^{12}$C, $^{14}$N and
$^{16}$O. The $\gamma $-decay corrected temperature $T_{\gamma}$ is
found to be insensitive to the choice of fragment set 
and underestimates $T_{fz} ( = T_{\gamma +p+sc})$ considerably.
Inclusion of particle-decay narrows the gap from $T_{fz}$ significantly,
particularly for the broader set of fragment species.
Right panels of Fig.~2 display 
the extracted volume $V_{app}$  as a function
of the freeze-out temperature.
The $\gamma$-decay corrected volume $V_{\gamma}$ overestimates
$V_{fz}$ significantly. Inclusion of particle-decay brings it closer
to $V_{fz}$, particularly for the larger set. The uncertainty in
the $(\gamma +p)$-corrected value for the volume, with $n_{iso}$=13, is
seen to be at most 25{\%} and that for temperature, it is at most
5{\%}. Inclusion of heavier species in the fitting procedure masks
the scattering effects. The calculations have been repeated for
$V_{fz}$=$6V_0$ and $8V_0$; the conclusions do not change for this
range of freeze-out volumes.

Along with temperature and volume, the nucleon chemical potentials
$\mu_n$ and $\mu_p$ are also extracted 
in this method which are not shown here. With
the knowledge of these four thermodynamic parameters, it is
straightforward to determine the entropy of the disassembling
system. Thus the evolution of entropy with
$T_{fz}$ can be known which acts as an important 
signature for the liquid-gas type
phase transition. This will be reported elsewhere.

In this paper, limitations of the currently used isotope thermometry
to determine the temperature and volume of a hot fragmenting nuclear
system has been pointed out. It is stressed that the strong interaction
effects left out in such a determination leaves a sizeable
uncertainty. This has an important bearing on many predictions on the
properties of hot finite nuclear matter. A new method, namely the 
least-squares fit to the fragment multiplicities is proposed to
extract the thermodynamic observables. 
We find this more promising in a numerical experiment,
this can be readily implemented in a realistic
experimental situation.\\

The authors acknowledge the support from the Department
of Science  \& Technology, Government of India.

\end{document}